\documentclass[aps,twocolumn,prm,superscriptaddress,longbibliography]{revtex4-1}
\usepackage{times,amsmath,amssymb,graphicx,multirow,booktabs}
\usepackage[colorlinks=true,citecolor=blue,urlcolor=blue,anchorcolor=blue,linkcolor=blue]{hyperref}



\begin{document}
\title{Deep-learning Hamiltonian reveals twist-tunable flat bands and nonlinear photocurrents in SrTiO$_3$ moir\'e bilayers}

\author{Meiyang Yu}
\affiliation{State Key Laboratory of Integrated Optoelectronics, Key Laboratory of Automobile Materials of MOE, and School of Materials Science and Engineering, Jilin University, Changchun, China}
\affiliation{Suzhou Laboratory, Suzhou, China}

\author{Chen Shen}
\email{shenc@szlab.ac.cn}
\affiliation{Suzhou Laboratory, Suzhou, China}

\author{Ruiwen Xie}
\email{xie.ruiwen@hotmail.com}
\affiliation{Institute of Materials Science, Technische Universität Darmstadt, Darmstadt, Germany
}

\author{Jingwei Tao}
\affiliation{State Key Laboratory of Integrated Optoelectronics, Key Laboratory of Automobile Materials of MOE, and School of Materials Science and Engineering, Jilin University, Changchun, China}

\author{Lijun Zhang}
\affiliation{State Key Laboratory of Integrated Optoelectronics, Key Laboratory of Automobile Materials of MOE, and School of Materials Science and Engineering, Jilin University, Changchun, China}

\author{Hongbin Zhang}
\email{hzhang@tmm.tu-darmstadt.de}
\affiliation{Institute of Materials Science, Technische Universität Darmstadt, Darmstadt, Germany
}

\date{\today}

\begin{abstract}
The extension of moir'e physics to complex oxides offers new ways to manipulate electronic states, but the large oxide moir'e supercells make systematic first-principles calculations demanding. Here, we combine density functional theory with the E(3)-equivariant deep-learning Hamiltonian framework DeepH-E3 to investigate the twist-angle-dependent electronic structure and optical responses of twisted bilayer SrTiO$_3$.  The model is trained on untwisted bilayers with different interlayer-sliding configurations and then applied to commensurate twisted bilayers with twist angles from $8.80^\circ$ to $53.13^\circ$. Compared with the untwisted bilayer, decreasing twist angle systematically flattens the valence bands and leads to nearly dispersionless bands at the smallest angles studied. Based on the predicted Hamiltonians, we evaluate the dielectric response, second-harmonic generation (SHG), shift current, and spin Hall conductivity. The dielectric response and spin Hall conductivity remain close to those of the untwisted bilayer, whereas the nonlinear optical responses are more strongly affected by twisting. SHG is strongly enhanced relative to the weak untwisted response, and the shift current shows a clear twist-angle dependence within the response-calculation range ($53.13^\circ$--$22.62^\circ$). These results show that twist engineering can control electronic and optoelectronic responses in oxide moir\'e systems.
\end{abstract}

\maketitle

\section{Introduction}

Twist-angle control has become an important way to modify the electronic structure of layered materials\cite{Carr2020,Andrei2020}. In magic-angle twisted bilayer graphene (TBG), a long-wavelength moir\'e potential produces nearly flat bands near $\theta = 1.1^\circ$\cite{Bistritzer2011,1Cao2018,Cao2018}. The reduced bandwidth enhances the role of electron-electron interactions\cite{Choi2019,Lee2019}. As a result, correlated insulating states\cite{1Cao2018}, unconventional superconductivity\cite{Cao2018,Yankowitz2019,Hao2021}, topological phases\cite{Choi2021,Nuckolls2020,Polshyn2020}, and quantum anomalous Hall behavior\cite{Serlin2020} can appear within a narrow energy window. The success of TBG has motivated the extension of twist-angle engineering to other van der Waals materials, including transition-metal dichalcogenides (TMDs, e.g., MoTe$_2$ and WSe$_2$)\cite{Wang2020,Xian2021,Devakul2021,Xu2020}, hexagonal boron nitride (h-BN)\cite{Xian2019,Walet2021,Su2022}, and InSe\cite{Li2021}. In these systems, moir\'e excitons, topological flat bands, and correlated insulating states have also been reported. More generally, twisting introduces a spatially varying stacking environment and modifies the moir\'e period and interlayer coupling, thereby tuning the bandwidth, symmetry, and localization of low-energy electronic states\cite{Carr2020,Mak2022}.

Complex oxides provide a different setting for twist engineering because their physical properties are governed by coupled lattice, orbital, charge, and spin degrees of freedom. This coupling makes correlated oxides highly responsive to electric fields, magnetic fields, epitaxial strain, and interface geometry\cite{Hwang2012,Tokura2000}. Twisting provides another perturbation, but one that acts through a spatially periodic modulation of local stacking and interlayer coupling. Recent progress in freestanding perovskite membranes has made twisted oxide bilayers experimentally feasible\cite{Ji2019}. In ferroelectric BaTiO$_3$, twisting can generate polar vortex textures, local dipole modulations, and quasi-flat valence bands associated with Lieb-lattice-like localization\cite{Lee2024}. SrTiO$_3$ is a related but distinct perovskite oxide. It is a quantum paraelectric whose low-energy electronic structure is mainly formed by Ti 3d and O 2p orbitals. Recent studies and reviews of LaAlO$_3$/SrTiO$_3$ and related SrTiO$_3$-based heterointerfaces show that two-dimensional electron gases, superconductivity, and spin-orbit effects are strongly affected by stoichiometry, electrostatic gating, and broken interfacial symmetry\cite{Chen2024,Singh2024}. These results indicate that SrTiO$_3$ is sensitive to interface termination and local electrostatic environment. Recent theoretical and experimental work further suggests that twisted SrTiO$_3$ bilayers can host moir\'e-induced structural reconstruction and flat-band-like electronic features\cite{Shahed2025,Zhang2025}. An independent preprint on twisted SrTiO$_3$ membrane bilayers has also reported interface-sensitive Raman modes and second-harmonic generation near $\theta\simeq36^\circ$\cite{https://doi.org/10.48550/arxiv.2606.06289}. However, a systematic description of twist-angle-dependent electronic structure and optical response in bilayer SrTiO$_3$ is still lacking.

Twisted moir\'e systems present a major challenge for electronic-structure calculations because the moir\'e supercell grows rapidly at small twist angles. The computational cost of density functional theory (DFT) therefore increases steeply with system size, making systematic calculations across twist angle difficult. Machine-learning Hamiltonian methods provide an efficient route to this bottleneck. The DeepH framework learns DFT Hamiltonian matrix elements directly from first-principles data, and its E(3)-equivariant extension, DeepH-E3, incorporates crystalline symmetries into the neural-network architecture\cite{Li2022,Gong2023}. These approaches have demonstrated near first-principles accuracy in twisted graphene and van der Waals moir\'e materials, including the description of flat-band evolution and spin-orbit-coupling-driven band reconstruction\cite{Gong2023,Yang2024}. Recent developments have further extended DeepH-based workflows to plane-wave DFT data, hybrid-functional Hamiltonians, and perturbative response calculations\cite{Gong2024,Tang2024,Li2024}. Applying such approaches to twisted oxides remains challenging because Ti--O bonding, interlayer coupling, and local stacking variations must be described within the same Hamiltonian\cite{Lee2019,Liu2014}. Because lattice, charge, spin, and orbital degrees of freedom are strongly coupled in perovskite transition-metal oxides, these systems are difficult to reduce to continuum or few-parameter tight-binding models\cite{Bistritzer2011,Liu2014,Shahed2025}. This complexity, together with their stacking sensitivity, makes twisted oxide bilayers a natural target for deep-learning Hamiltonian methods, while a full treatment of all coupled responses remains an open direction.

In this work, we investigate the twist-angle dependence of the electronic structure and optical responses of bilayer SrTiO$_3$ using DeepH-E3. The model is constructed and validated using untwisted bilayer structures with different interlayer-sliding configurations, and then applied to commensurate twisted bilayer SrTiO$_3$ systems with twist angles from $53.13^\circ$ to $8.80^\circ$. By comparing the twisted systems with the untwisted bilayer, we reveal significant band reconstruction and the emergence of nearly flat valence bands as the twist angle decreases. Based on the predicted Hamiltonians, we further calculate the dielectric response, second-harmonic generation, shift current, and spin Hall conductivity and examine their variation with twist angle relative to the untwisted bilayer reference. Our results show that twisting modifies both the electronic structure and optical responses of bilayer SrTiO$_3$, with a particularly strong effect on nonlinear optical responses, including second-harmonic generation and the shift current.

\section{Computational Methods}
\begin{figure}[htbp]
	\centering
	\includegraphics[width=1\linewidth]{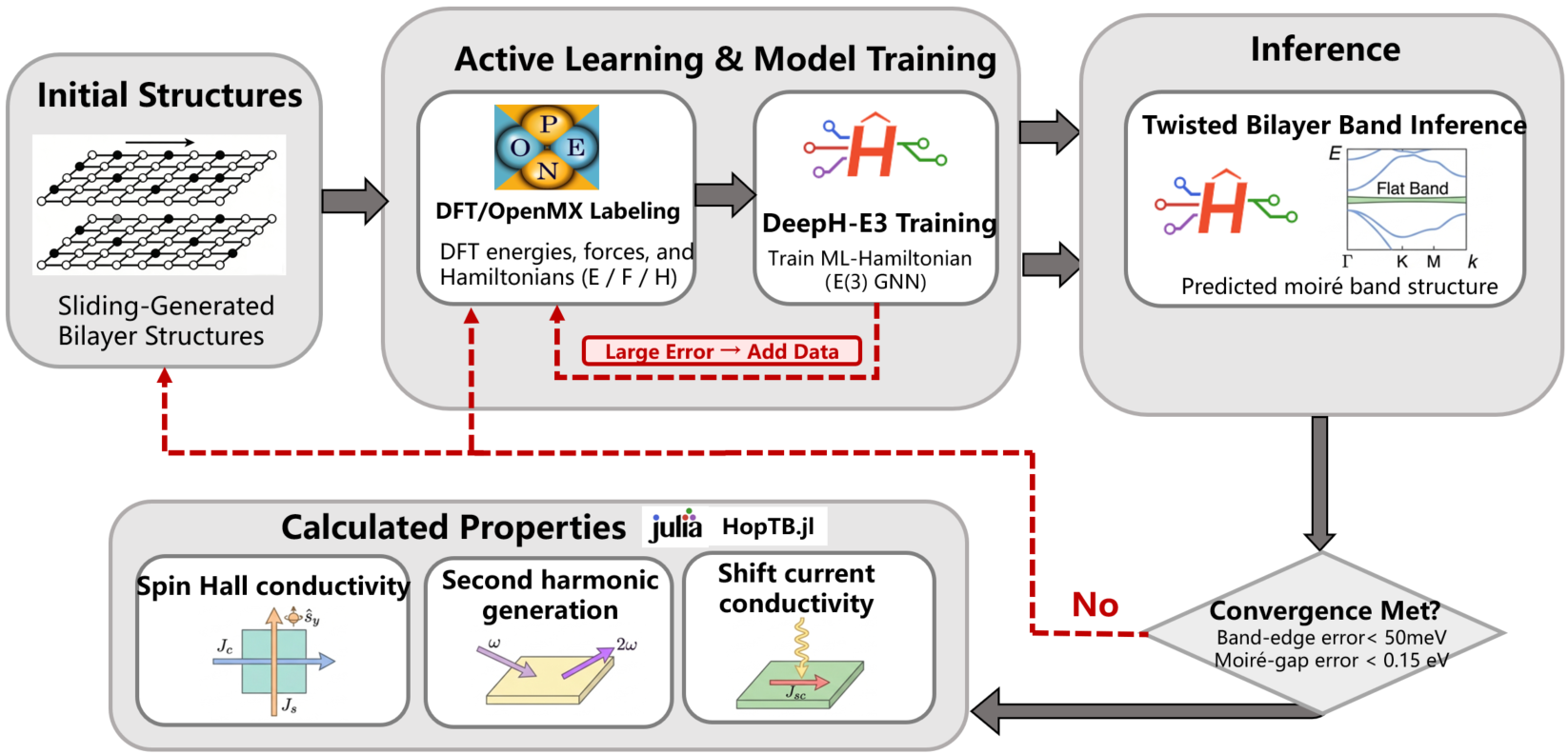}
	\caption{Deep-learning Hamiltonian workflow for twisted bilayer SrTiO$_3$. Sliding untwisted bilayers provide local stacking environments for OpenMX Hamiltonian generation and DeepH-E3 training. The trained model predicts Hamiltonians for commensurate twisted bilayers, which are then used to compute band structures and response functions with HopTB.}
	\label{fgr:fig-1}
\end{figure}

The overall computational workflow is illustrated in Fig.~\ref{fgr:fig-1}. All first-principles calculations were performed using the OpenMX package~\cite{Ozaki2003,Ozaki2004}, employing norm-conserving pseudopotentials and pseudoatomic localized basis functions. The exchange--correlation interaction was treated within the generalized gradient approximation in the Perdew--Burke--Ernzerhof (PBE) form~\cite{perdew1996generalized}. For Sr, Ti, and O atoms, the basis sets Sr10.0-s2p2d1, Ti7.0-s2p2d1, and O6.0-s2p2d1 were adopted, corresponding to cutoff radii of 10.0, 7.0, and 6.0 Bohr, respectively. These basis sets include the valence orbitals Sr($5s,5p,4d$), Ti($4s,4p,3d$), and O($2s,2p,3d$). Spin--orbit coupling was included in the noncollinear formalism, using a two-component spinor basis with 13 spatial orbitals (26 spinor components) per Sr, Ti, and O site in the Hamiltonian.
The untwisted bilayer $3 \times 3$ supercells containing 126 atoms were used to construct the training dataset for the stacking-dependent Hamiltonian, for which Monkhorst--Pack $k$-point meshes determined from convergence tests were employed. A total of 623 sliding configurations with different interlayer translations were generated and randomly divided into training, validation, and test sets using a 3:1:1 ratio. A real-space grid cutoff of 400~Ry was adopted. Electronic occupations were treated using a Fermi--Dirac smearing corresponding to 300~K, and the self-consistent-field convergence criterion was set to $4 \times 10^{-8}$~Hartree. The resulting first-principles calculations provide the Hamiltonian matrices $H_{i\alpha,j\beta}$ in a localized-orbital basis, which serve as reference data for model construction. 

The mapping from atomic structures to Hamiltonian elements is established using an E(3)-equivariant graph neural network (DeepH)\cite{Li2022,Gong2023}, which predicts localized-orbital Hamiltonian elements $H_{i\alpha,j\beta}$ while preserving spatial symmetries and incorporating spin--orbit coupling. The model is trained by minimizing the reconstruction error with respect to first-principles Hamiltonians, using the Adam optimizer with an initial learning rate of $1.25 \times 10^{-3}$ and a small batch size due to the large size of individual structures. An active learning scheme is employed to iteratively improve model transferability. The trained model was subsequently applied to twisted bilayer supercells with much larger atom numbers, for which only the $\Gamma$ point was used for Brillouin-zone sampling.

After convergence, the learned Hamiltonians were used to reconstruct the electronic band structures and evaluate electronic and optical response functions. The predicted Hamiltonians were interfaced with the HopTB package~\cite{HopTB,Wang2017,Wang2019}, which implements linear-response, nonlinear-optical, and spin-transport formalisms within a tight-binding framework. The dielectric function, second-harmonic generation (SHG), shift current, and intrinsic spin Hall conductivity (SHC) were calculated using the corresponding implementations of Wang \textit{et al.}~\cite{Wang2017,Wang2019}.
A uniform broadening parameter of $\eta=0.1$~eV was adopted to account for finite lifetime effects in the response calculations. Brillouin-zone integrations were performed using converged $k$-point meshes for both untwisted and twisted bilayer structures. Convergence tests for the broadening parameter and Brillouin-zone sampling are provided in the Supplemental Material (Figs.~S8--S10). Unless otherwise stated, all optical and transport properties were evaluated using the corresponding converged parameters.

\section{Results and Discussion}

\subsection{Twisted bilayer structures and training dataset}

\begin{figure}[htbp]
	\centering
	\includegraphics[width=1\linewidth]{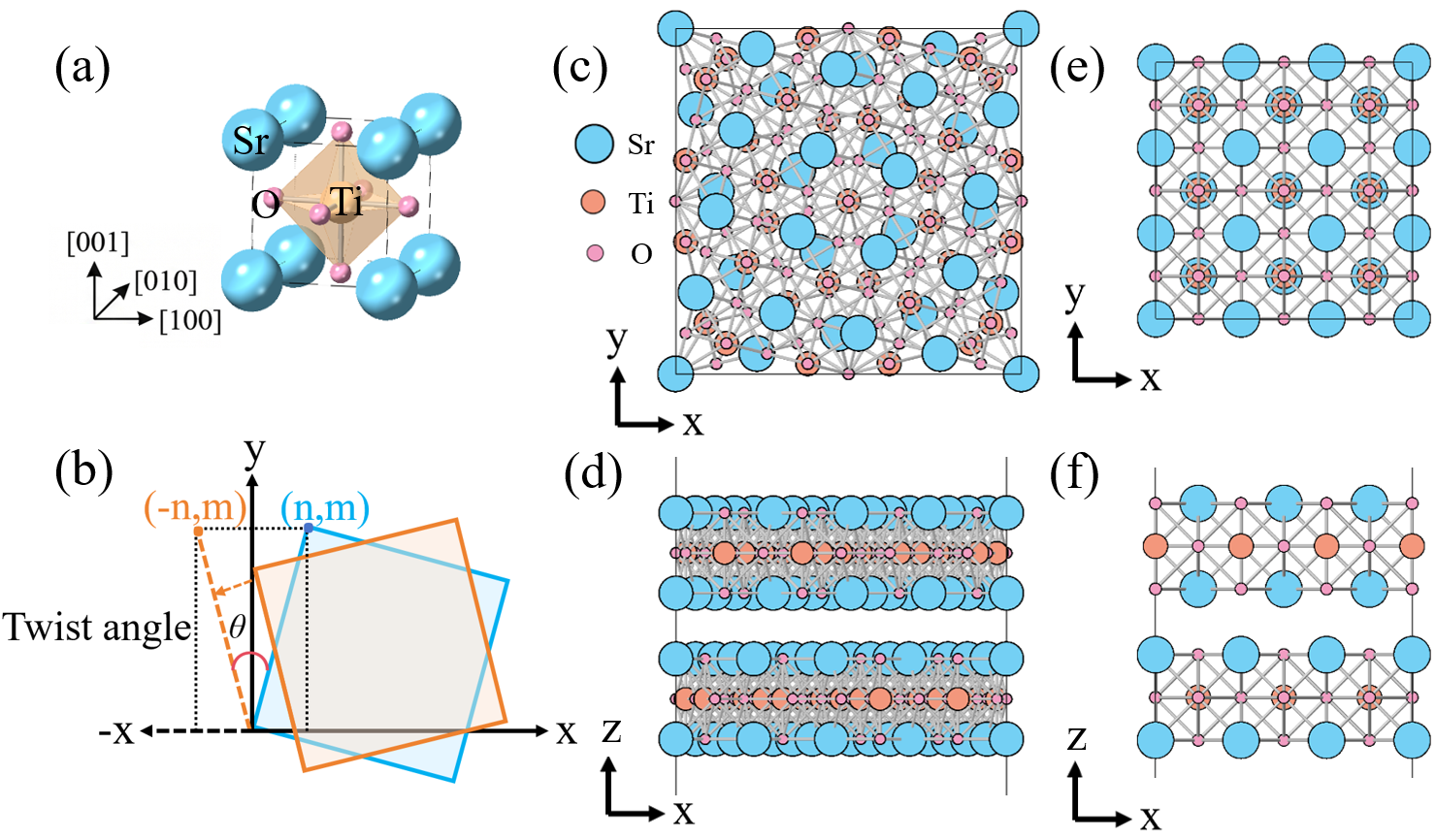}
	\caption{
Dataset construction and structural configurations of bilayer SrTiO$_3$.
(a) Conventional cubic unit cell of SrTiO$_3$.
(b) Schematic illustration of the twist operation.
(c,d) Top and side views of a representative twisted bilayer with $\theta = 28.07^\circ$, corresponding to the $(1,4)$ supercell.
(e,f) Top and side views of representative sliding bilayer configurations used for Hamiltonian learning, generated by in-plane translations between the two layers.
}
	\label{fgr:fig-2}
\end{figure}

We first describe the structural models used in this work, including the twisted bilayer SrTiO$_3$ systems investigated in the following sections and the sliding bilayer configurations employed for Hamiltonian learning. The twisted structures serve as the target moir\'e systems, whereas the sliding configurations sample the local stacking environments needed to learn the effective Hamiltonian. Strontium titanate (SrTiO$_3$) is a cubic perovskite oxide with space group \textit{Pm$\bar{3}$m} and lattice constant $a \approx 3.905$~\AA, as shown in Fig.~\ref{fgr:fig-2}(a). The structure consists of alternating SrO and TiO$_2$ layers stacked along the [001] direction, where Ti atoms are octahedrally coordinated by oxygen to form TiO$_6$ units.

To investigate twist-induced effects, we construct twisted bilayer SrTiO$_3$ based on commensurate supercells under periodic boundary conditions. As illustrated in Fig.~\ref{fgr:fig-2}(b), the twist is introduced as a relative rotation between the two layers around the out-of-plane axis. The twist angle $\theta$ is defined by integer pairs $(n,m)$ that determine the superlattice vectors, following standard constructions for twisted bilayer systems~\cite{Mele2010}, leading to the relation $\theta = 2\arctan(n/m)$. This relative rotation gives rise to a moir\'e superlattice characterized by a long-wavelength modulation of local atomic registries. The resulting moir\'e potential modifies the local stacking environment and can reshape the low-energy electronic structure~\cite{Mak2022,Carr2020}. All bilayer structures are constructed using a symmetric SrO--SrO interface with a fixed interlayer distance of 3.2~\AA, consistent with previous studies on twisted perovskite systems such as bilayer BaTiO$_3$~\cite{Lee2024}. The number of atoms in the supercell is determined by $(n,m)$; the largest supercell considered in this work, (1,13), contains 2380 atoms, making conventional DFT calculations impractical and motivating the machine-learning approach. A representative twisted structure is shown in Figs.~\ref{fgr:fig-2}(c) and \ref{fgr:fig-2}(d), corresponding to the $(1,4)$ supercell with $\theta = 28.07^\circ$ and 408 atoms, which exhibits a clear moir\'e superlattice.

To capture the local stacking variations present in the moir\'e superlattices, we constructed a set of sliding bilayer configurations by translating the top layer relative to the bottom layer. Representative structures are shown in Figs.~\ref{fgr:fig-2}(e) and \ref{fgr:fig-2}(f). These structures cover a wide range of local stacking registries and provide the training basis for learning the moir\'e Hamiltonian. Details of the dataset construction, including the total number of configurations and the training, validation, and test split, are provided in the Supplemental Material.

\subsection{Deep Hamiltonian validation}

\begin{figure*}[t]
    \centering
    \includegraphics[width=0.85\textwidth]{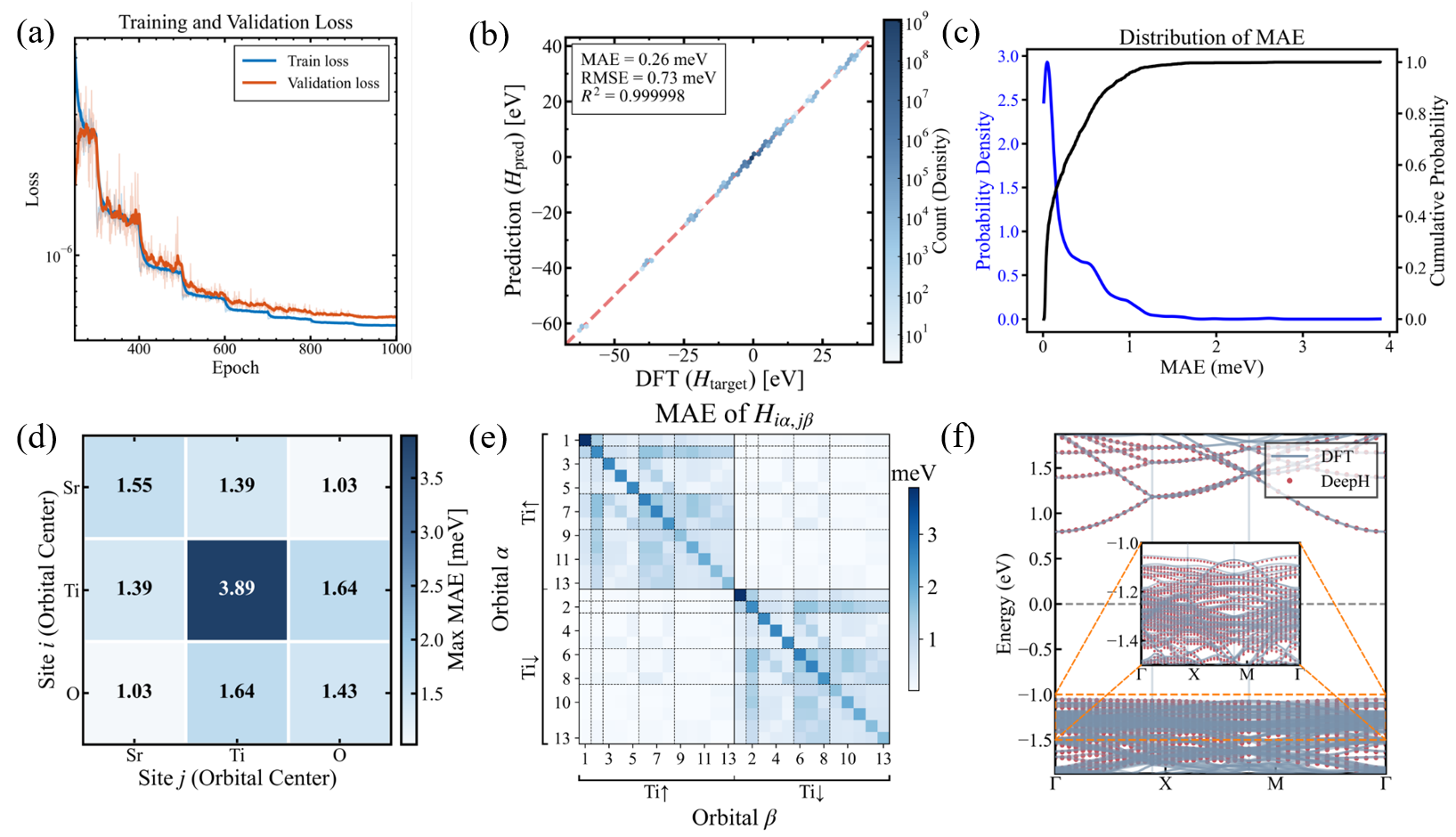}
    \caption{Validation of the DeepH-E3 model for twisted SrTiO$_3$.
    (a) Training and validation losses.
    (b) Parity plot of DeepH-E3-predicted and DFT-calculated Hamiltonian matrix elements.
    (c) Probability density function (PDF) and cumulative distribution function (CDF) of the mean absolute errors (MAEs).
    (d) Element-resolved MAE of the Hamiltonian matrix elements $H_{i\alpha,j\beta}$.
    (e) Orbital-resolved MAE of the $13\times13$ Ti--Ti hopping block.
    (f) Comparison of DFT (solid lines) and DeepH-E3 (dashed lines) band structures for twisted SrTiO$_3$ ($\theta=28.07^\circ$). The inset shows the valence-band maximum region.}
    \label{fgr:fig-3}
\end{figure*}

The sliding bilayer dataset described above is used to train the DeepH-E3 model. We assess the learned Hamiltonian using training convergence, statistical error metrics, and band-structure comparisons with DFT. As shown in Fig.~\ref{fgr:fig-3}(a), the training and validation losses decrease rapidly during the early stages of training and converge after approximately 700 epochs. The validation loss closely follows the training loss throughout the optimization process and does not exhibit any systematic increase at later epochs, indicating no apparent overfitting.

The predictive accuracy of the Hamiltonian matrix elements is further quantified in Figs.~\ref{fgr:fig-3}(b) and \ref{fgr:fig-3}(c). Fig.~\ref{fgr:fig-3}(b) shows the correlation between the predicted and DFT reference Hamiltonian matrix elements over an energy range approaching 100 eV. The reference matrix elements show a clustered and discontinuous distribution, which reflects the block structure of the Hamiltonian in a localized atomic-orbital basis\cite{Li2022,Gong2023}. This structure arises from the chosen basis set and leads to natural grouping of matrix elements into on-site and inter-site orbital channels\cite{Slater1954}. Across this broad and discrete energy range, the predicted values closely follow the ideal diagonal, with a MAE of 0.26 meV, RMSE of 0.73 meV, and $R^2 = 0.999998$. The error distribution, shown in Fig.~\ref{fgr:fig-3}(c), exhibits a sharp peak in the sub-meV regime, indicating that most Hamiltonian elements are predicted with high accuracy. The cumulative distribution further shows that more than 99\% of the matrix elements have errors below 2 meV, while no error exceeds 4 meV.

The element-resolved maximum MAE of the Hamiltonian matrix elements is summarized in Fig.~\ref{fgr:fig-3}(d). All atomic interaction channels exhibit maximum errors below 4 meV. The largest deviation is found for the Ti--Ti interaction, with a maximum MAE of 3.89 meV, followed by 1.64 meV for the Ti--O interaction, whereas all remaining atomic pairs remain below 1.6 meV. The larger Ti-related errors are consistent with the stronger sensitivity of Ti-centered orbitals to local structural variations. Fig.~\ref{fgr:fig-3}(e) further presents the spin- and orbital-resolved MAE for the Ti-related matrix elements, corresponding to the Ti-centered 13$\times$13 orbital block. This block is constructed from the Ti($4s,4p,3d$) interaction channels. The corresponding spin- and orbital-resolved MAE distributions for the remaining interactions are shown in Fig.~S1 of the Supplemental Material. A clear block structure is observed, with larger errors primarily concentrated in the diagonal on-site terms, whereas the off-diagonal hopping elements exhibit smaller deviations. The spin-up and spin-down sectors exhibit nearly identical error distributions. These errors are much smaller than the characteristic energy scales of the Hamiltonian matrix elements, indicating that DeepH-E3 reproduces the DFT Hamiltonian with high fidelity.

Finally, the reliability of the model is validated at the level of electronic structure. As a representative validation case, Fig.~\ref{fgr:fig-3}(f) compares the DeepH-E3 and DFT band structures for the $(1,4)$ twisted bilayer with $\theta = 28.07^\circ$. The band structure obtained from the DeepH-E3 Hamiltonian shows excellent agreement with the DFT reference along high-symmetry paths. The overall dispersions, band edges, and characteristic features are accurately reproduced. The agreement remains excellent even in regions containing densely entangled bands, as highlighted in the inset. In particular, fine structures associated with the moir\'e superlattice, such as flat bands and band splittings, are well captured. Additional band-structure validations at other twist angles and the computational-cost comparison between the DFT and DeepH-E3 workflows are provided in Fig.~S2 of the Supplemental Material. Together with previous DeepH studies\cite{Li2022,Gong2023}, these results support the use of the DeepH-E3 Hamiltonian for large-scale simulations of twisted SrTiO$_3$ moir\'e superlattices.

\subsection{Evolution of moir\'e electronic structures}

\begin{figure*}[t]
    \centering
    \includegraphics[width=0.81\textwidth]{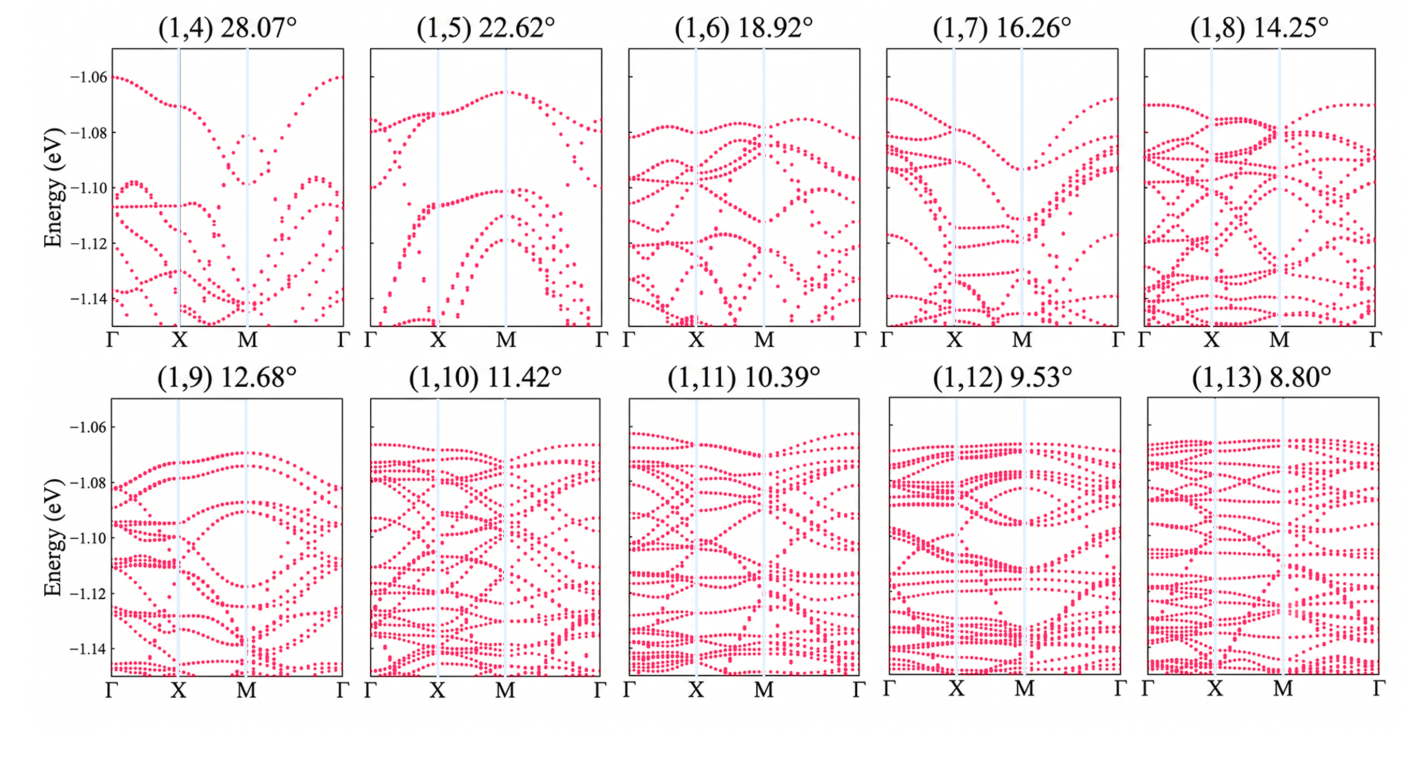}
	\caption{Twist-angle-dependent VBM-region band structures of bilayer SrTiO$_3$ predicted by the DeepH-E3 model for ten representative commensurate configurations ranging from $(1,4)$ to $(1,13)$. 
The band structures are calculated along the high-symmetry path $\Gamma$--X--M--$\Gamma$. 
As the twist angle decreases, the bands flatten and form nearly dispersionless flat bands in the small-angle regime.
}
	\label{fgr:fig-4}
\end{figure*}

\begin{figure}[t]
    \centering
    \includegraphics[width=\linewidth]{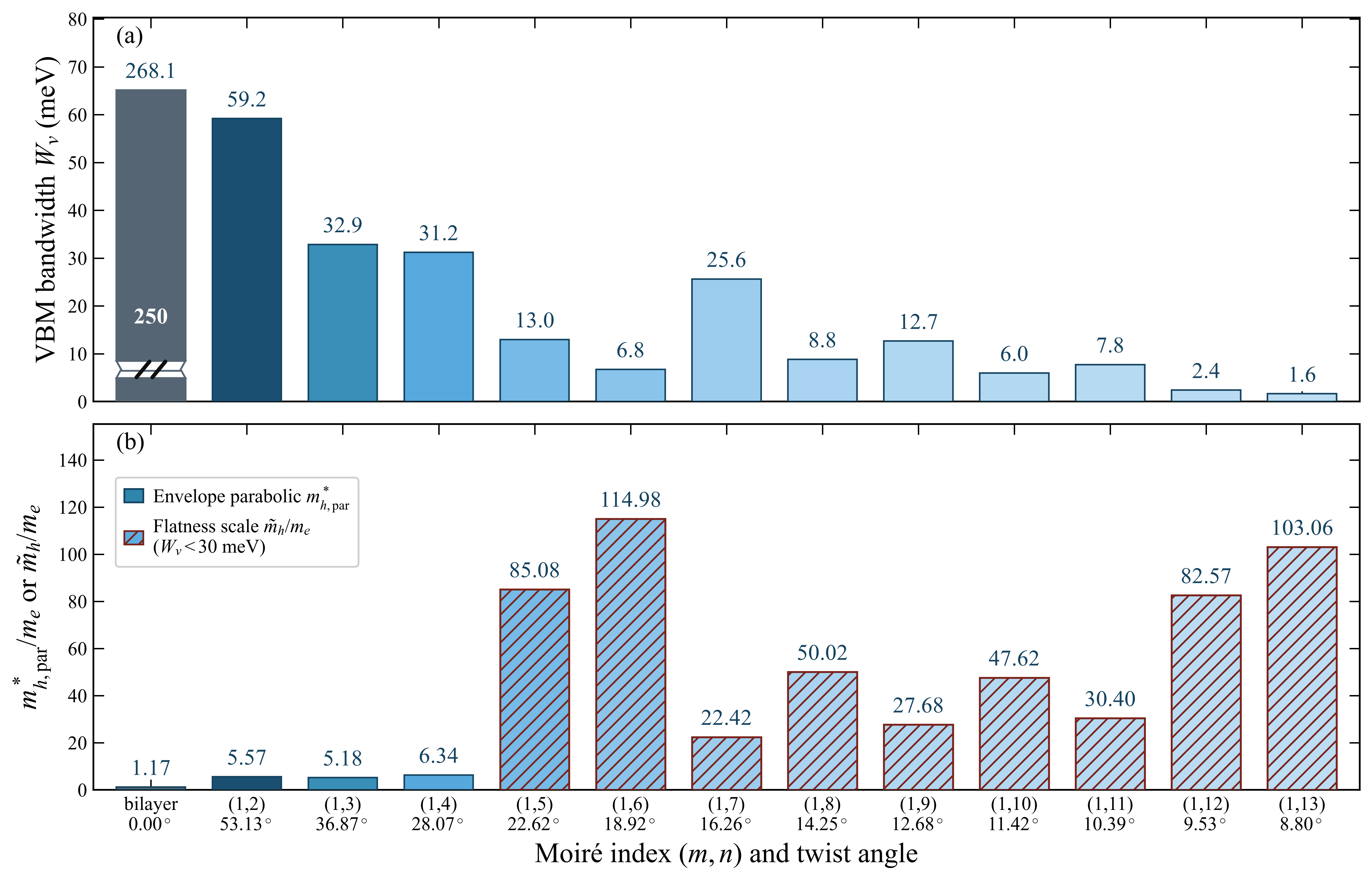} 
\caption{Evolution of valence-band bandwidth and reported hole mass with twist angle.
(a) VBM bandwidth $W_v$ for representative moir\'e configurations $(m,n)$.
(b) Reported hole mass: $m_{h,\mathrm{par}}^*/m_e$ from a seven-point envelope fit ($W_v \ge 30$~meV)
or $\tilde{m}_h/m_e$ from the bandwidth-derived flatness scale ($W_v < 30$~meV).
Hatched bars mark $\tilde{m}_h$.}
    \label{fgr:fig-5}
\end{figure}

With the DeepH-E3 model validated, we examine the twist-angle-dependent electronic structure of bilayer SrTiO$_3$. Figure~\ref{fgr:fig-4} shows the VBM-region band structures for ten commensurate configurations from $(1,4)$ to $(1,13)$ ($28.07^\circ$ to $8.80^\circ$), plotted along the $\Gamma$--X--M--$\Gamma$ path within the same energy window for direct comparison. The complete band structures, including the CBM region and the untwisted, $(1,2)$, and $(1,3)$ configurations, are provided in Fig.~S3 of the Supplemental Material, where the CBM shows weaker flattening than the VBM. Relative to the untwisted bilayer, the VBM bands show clear moir\'e-induced reconstruction as the twist angle decreases. In the large-angle regime ($\theta > 20^\circ$), the VBM remains relatively dispersive, with only weak band folding and small energy splittings induced by the moir\'e superlattice. As the twist angle is reduced to the intermediate regime ($15^\circ \lesssim \theta \lesssim 20^\circ$), the shrinking moir\'e Brillouin zone gives rise to more subbands and a visible reduction of the overall dispersion~\cite{Bistritzer2011,Cao2018}. In the smaller-angle structures, several VBM branches become nearly flat near the band edge, consistent with a strong reduction of the hole kinetic energy. Although the evolution is not perfectly monotonic from one commensurate cell to the next, the overall reduction of VBM dispersion with decreasing twist angle is clear.

To quantify this trend, we analyze the VBM bandwidth and reported hole mass in Fig.~\ref{fgr:fig-5}. Because a local parabolic curvature becomes poorly defined once the VBM dispersion is reduced to the meV scale, we use a path-resolved parabolic fit for dispersive bands and a bandwidth-derived flatness scale for nearly flat VBMs. From the VBM envelope $E_v(s)$ along $\Gamma$--$X$--$M$--$\Gamma$, the quantity plotted in Fig.~\ref{fgr:fig-5}(b) is obtained as
\begin{equation}
  \frac{m_{h,\mathrm{par}}^*}{m_e}
  = \frac{\hbar^2/m_e}{\left|d^2E_v/ds^2\right|}
  \quad (W_v \ge 30~\mathrm{meV}),
\end{equation}
using a seven-point parabolic fit at the VBM maximum, and as
\begin{equation}
  \frac{\tilde{m}_h}{m_e}
  = \frac{\hbar^2/m_e}{2W_v/(\Delta s)^2}
  \quad (W_v < 30~\mathrm{meV}),
\end{equation}
where $\Delta s$ is the total cumulative path length and hatched bars mark $\tilde{m}_h$. As shown in Fig.~\ref{fgr:fig-5}(a), the VBM bandwidth $W_v$ is strongly reduced in most twisted structures compared with the untwisted bilayer. Most configurations from $(1,5)$ onward satisfy $W_v < 30$~meV, and the smallest-angle structures reach bandwidths of only a few meV, confirming the formation of nearly flat VBM minibands. The corresponding reported mass in Fig.~\ref{fgr:fig-5}(b) increases markedly in the flat-band regime, consistent with the suppression of hole kinetic energy. At the smallest twist angles, $\tilde{m}_h/m_e$ reaches values of roughly 22--115, more than an order of magnitude larger than in the untwisted bilayer. The remaining non-monotonic variations in both $W_v$ and the reported mass ($m_{h,\mathrm{par}}^*$ or $\tilde{m}_h$) indicate that the VBM minibands are sensitive to the discrete commensurate stacking geometry, rather than being determined by twist angle alone~\cite{Naik2018,Shahed2025}. Additional band structures, bandwidth analysis, path-resolved mass fits, and CBM data are provided in Figs.~S3--S7 of the Supplemental Material.

The emergence of nearly flat VBM bands in twisted SrTiO$_3$ can be understood from the spatial variation of local stacking across the moir\'e supercell. Different stacking regions generate different electrostatic and crystal-field environments, which modify the Ti--O-derived band-edge states. This mechanism differs from that in graphene-based moir\'e systems, where flat bands are mainly associated with interlayer hopping interference~\cite{Bistritzer2011,Cao2018}. In the present oxide bilayer, the reduction of VBM dispersion is more naturally tied to stacking-dependent electrostatic and orbital hybridization effects~\cite{Zhang2025,Shahed2025}. This stacking sensitivity also explains the non-monotonic variations in $W_v$ and the reported mass: each commensurate structure samples the local stacking landscape differently, so the VBM curvature and bandwidth need not vary smoothly with twist angle. These results show that twist angle can tune the low-energy electronic structure of bilayer SrTiO$_3$, while also highlighting the sensitivity of oxide moir\'e bands to the discrete commensurate geometry.

\subsection{Electronic and nonlinear optical responses}

\begin{figure*}[htbp]
	\centering
	\includegraphics[width=0.75\linewidth]{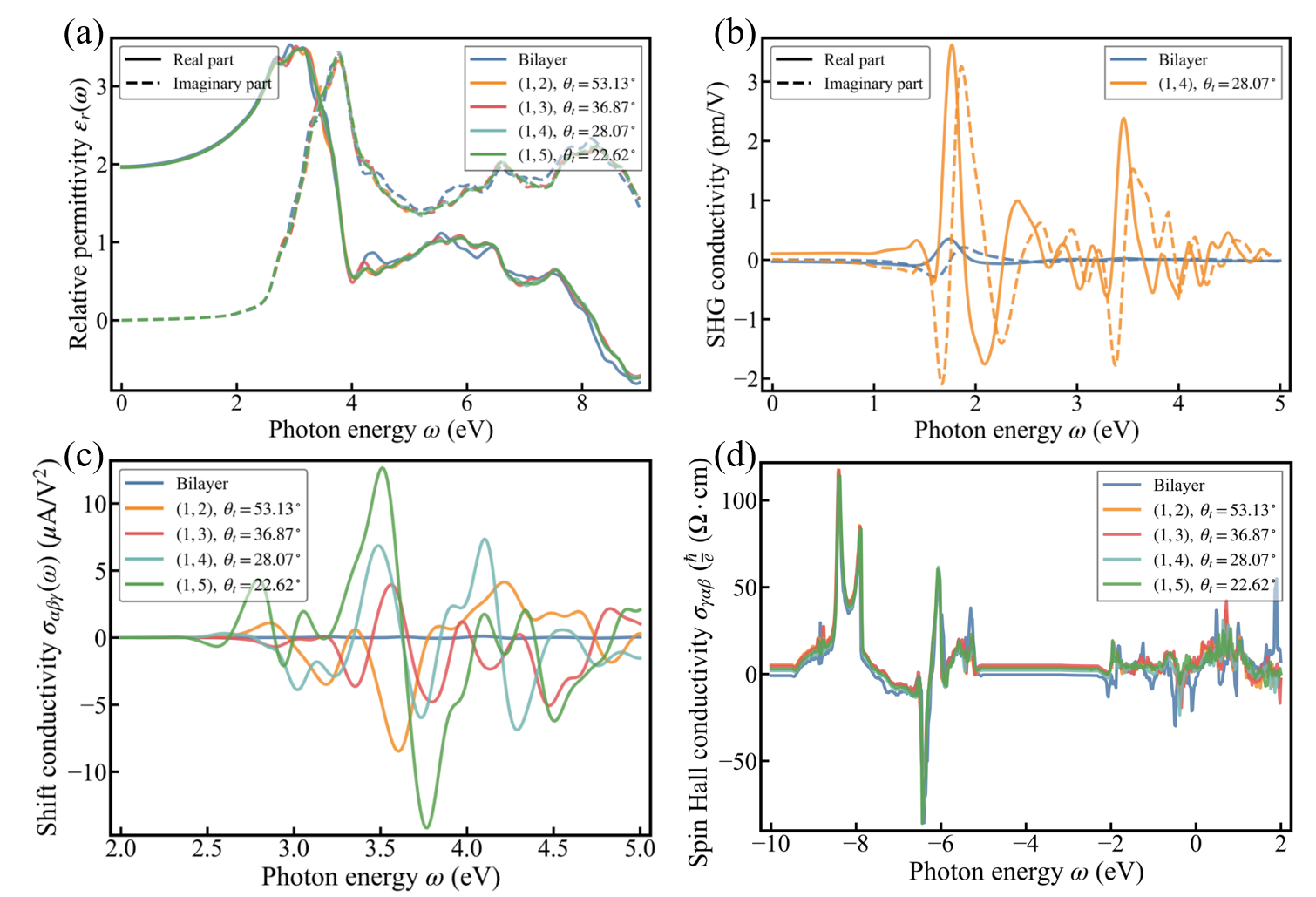}
	\caption{Evolution of optical and transport responses from untwisted to twisted bilayer SrTiO$_3$. (a) Dielectric functions of the untwisted bilayer and moir\'e configurations with different twist angles; solid and dashed lines denote the real and imaginary parts, respectively. (b) SHG responses of the untwisted bilayer and the $(1,4)$ structure with $\theta = 28.07^\circ$. (c) Shift current responses as a function of photon energy for moir\'e configurations with different twist angles. (d) SHC as a function of chemical potential for the same twist angles.}
	\label{fgr:fig-6}
\end{figure*}

Using the Hamiltonians predicted by the DeepH-E3 model, we calculate the dielectric response, SHG, shift current, and SHC for representative moir\'e configurations at large-to-intermediate twist angles ($53.13^\circ$, $36.87^\circ$, $28.07^\circ$, and $22.62^\circ$), with SHG compared between the untwisted bilayer and the $(1,4)$ structure. As shown in Fig.~\ref{fgr:fig-6}, these responses exhibit clear twist-angle dependence within this range.

The imaginary part, $\mathrm{Im}[\varepsilon_{r}]$, describes interband optical absorption, while the real part, $\mathrm{Re}[\varepsilon_{r}]$, characterizes the corresponding dispersive response. Fig.~\ref{fgr:fig-6}(a) shows the calculated dielectric function for the untwisted bilayer and representative moir\'e configurations. The spectrum shows the characteristic response of a wide-gap oxide insulator: $\mathrm{Im}[\varepsilon_{r}]$ remains nearly zero below an optical gap of approximately $2.5$\,eV and develops a broad absorption peak around $3.8$\,eV, consistent with interband transitions between O $2p$ valence states and Ti $3d$ conduction states. A secondary absorption feature appears near $\sim6.5$\,eV. Correspondingly, $\mathrm{Re}[\varepsilon_{r}]$ increases from a static value of approximately $3.0$ to a broad maximum near $3.5$\,eV, followed by a dispersive reduction at higher photon energies. Over the calculated angle range, the dielectric spectra closely overlap with the untwisted-bilayer response. The optical gap, static dielectric constant, and main absorption features vary only weakly with twist angle, indicating that the linear optical response is weakly affected by the moir\'e modulation in this large-to-intermediate angle range.

We then turn to SHG, a second-order nonlinear optical response sensitive to moir\'e-induced inversion-symmetry breaking. We focus on the in-plane $\sigma^{xxx}$ component. Fig.~\ref{fgr:fig-6}(b) shows $\mathrm{Re}[\sigma^{xxx}(2\omega)]$ and $\mathrm{Im}[\sigma^{xxx}(2\omega)]$ for the untwisted bilayer and the $(1,4)$ structure ($\theta=28.07^{\circ}$). The response remains negligible below $\sim1.5$\,eV and increases at higher photon energies, with a main resonance near $1.8$\,eV and a broader feature in the $3$--$4$\,eV range. The dominant SHG peak near $\sim1.8$\,eV corresponds to a two-photon excitation energy close to the main absorption feature in $\mathrm{Im}[\varepsilon_r]$, suggesting an association with interband transitions between the valence-band edge and the lowest conduction bands. The higher-energy SHG feature is similarly consistent with the secondary absorption structure in the dielectric spectrum. The untwisted bilayer gives a much weaker $\sigma^{xxx}$ than the twisted $(1,4)$ structure, showing that twisting strongly enhances SHG through interfacial inversion-symmetry breaking. A recent preprint on a twisted SrTiO$_3$ membrane bilayer at $\theta\simeq 36^\circ$ also reports a strong interfacial SHG signal~\cite{https://doi.org/10.48550/arxiv.2606.06289}. Our calculation likewise shows a finite SHG response in the $(1,4)$ model even though the VBM bands remain relatively dispersive at this angle.

We next examine the shift current, a second-order nonlinear photocurrent associated with the bulk photovoltaic effect. Compared with the linear dielectric response, the shift current is more sensitive to the underlying electronic structure and wave-function geometry. We focus on the in-plane $\sigma^{xxx}(\omega)$ component. Fig.~\ref{fgr:fig-6}(c) shows the calculated shift current for the untwisted bilayer and four commensurate moir\'e configurations with twist angles ranging from $53.13^\circ$ to $22.62^\circ$. The untwisted bilayer exhibits only a weak shift-current response, whereas the twisted structures are strongly enhanced. For the twisted systems, the response remains negligible below approximately $2.5~\mathrm{eV}$. At higher photon energies, all spectra show pronounced resonances, including a dominant positive peak around $3.0$--$3.8~\mathrm{eV}$ and a broader feature near $4~\mathrm{eV}$. The spectra also change sign with photon energy, indicating opposite photocurrent directions for different interband transitions~\cite{Sipe2000,Young2012}. Although the resonance energies vary only weakly with twist angle, the peak amplitudes increase markedly across the series. In particular, the main positive peak increases from approximately $4~\mu\mathrm{A}/\mathrm{V}^2$ in the $(1,2)$ configuration to about $12.5~\mu\mathrm{A}/\mathrm{V}^2$ in $(1,5)$, accompanied by a stronger negative resonance. This enhancement shows that the nonlinear photocurrent is more sensitive to the moir\'e perturbation than the linear dielectric response, even without pronounced flat bands or strong band reconstruction. The coexistence of positive and negative shift-current peaks indicates that different interband transitions drive photocurrents in opposite directions, suggesting a twist- and wavelength-dependent nonlinear photocurrent response~\cite{Spanier2016}.

Finally, we examine the SHC, which characterizes the generation of a transverse spin current under an applied electric field. In the insulating regime considered here, the SHC is dominated by the intrinsic contribution associated with the spin Berry curvature of the occupied states. We focus on the $\sigma_{\mathrm{SH}}^{zxy}$ component, corresponding to a $z$-polarized spin current flowing along $x$ in response to an electric field along $y$. Figure~\ref{fgr:fig-6}(d) shows the SHC as a function of chemical potential for the untwisted bilayer and the four moir\'e configurations. The SHC is concentrated mainly in the valence-band region between approximately $-10$ and $-2~\mathrm{eV}$, where several positive and negative resonance features appear. By contrast, it remains nearly zero within the band gap and in the conduction-band region, consistent with the insulating electronic structure. Across the calculated angle range, including the untwisted bilayer, the SHC peak positions and amplitudes remain nearly unchanged, apart from minor non-monotonic variations between commensurate configurations. The weak angle dependence suggests that the integrated spin Berry curvature is only weakly affected by the moir\'e modulation in the present angle range.

Taken together, these results show that the calculated responses differ markedly in their sensitivity to moir\'e modulation. The linear dielectric response remains nearly unchanged, indicating that the overall interband optical structure is only weakly modified by twisting in the calculated angle range. SHG and shift current are strongly enhanced when twisting breaks inversion symmetry at the interface, with the shift current showing the clearest twist-angle dependence. By contrast, the SHC remains nearly invariant, suggesting that the integrated spin Berry curvature is comparatively insensitive to the moir\'e perturbation in this large-to-intermediate angle regime.

\section{Conclusions}

In conclusion, we use a deep-Hamiltonian framework to investigate twisted bilayer SrTiO$_3$ across a broad range of commensurate twist angles. By comparing twisted structures with the untwisted bilayer reference, we find a systematic flattening of the valence bands with decreasing twist angle, accompanied by the emergence of nearly dispersionless flat bands in the small-angle regime. Based on the predicted Hamiltonians, we investigated the dielectric response, second-harmonic generation, shift current, and spin Hall conductivity. Twisting has only a weak effect on the dielectric response and spin Hall conductivity in the calculated large-to-intermediate angle range, but it enhances the nonlinear optical responses. In particular, SHG is strongly enhanced in the representative $(1,4)$ twisted structure, while the shift current shows the clearest twist-angle dependence and increases as the twist angle decreases from $53.13^\circ$ to $22.62^\circ$. These results show that twist angle provides an additional handle on the nonlinear optical responses of bilayer SrTiO$_3$, most prominently through the shift current. The present work also highlights the capability of deep-learning Hamiltonian approaches for large-scale electronic-structure calculations in complex twisted oxide materials.

\begin{acknowledgments}
The authors gratefully acknowledge the computing time provided to them on the high-performance computer Lichtenberg at the NHR Centers NHR4CES at TU Darmstadt. 
\end{acknowledgments}

\bibliography{bibl}

@article{Gong2023,
  title = {General framework for E(3)-equivariant neural network representation of density functional theory Hamiltonian},
  volume = {14},
  ISSN = {2041-1723},
  url = {http://dx.doi.org/10.1038/s41467-023-38468-8},
  DOI = {10.1038/s41467-023-38468-8},
  number = {1},
  journal = {Nature Communications},
  publisher = {Springer Science and Business Media LLC},
  author = {Gong,  Xiaoxun and Li,  He and Zou,  Nianlong and Xu,  Runzhang and Duan,  Wenhui and Xu,  Yong},
  year = {2023},
  month = May 
}

@article{Ozaki2003,
  title = {Variationally optimized atomic orbitals for large-scale electronic structures},
  volume = {67},
  ISSN = {1095-3795},
  url = {http://dx.doi.org/10.1103/PhysRevB.67.155108},
  DOI = {10.1103/physrevb.67.155108},
  number = {15},
  journal = {Physical Review B},
  publisher = {American Physical Society (APS)},
  author = {Ozaki,  T.},
  year = {2003},
  month = Apr 
}

@article{Li2022,
  title = {Deep-learning density functional theory Hamiltonian for efficient ab initio electronic-structure calculation},
  volume = {2},
  ISSN = {2662-8457},
  url = {http://dx.doi.org/10.1038/s43588-022-00265-6},
  DOI = {10.1038/s43588-022-00265-6},
  number = {6},
  journal = {Nature Computational Science},
  publisher = {Springer Science and Business Media LLC},
  author = {Li,  He and Wang,  Zun and Zou,  Nianlong and Ye,  Meng and Xu,  Runzhang and Gong,  Xiaoxun and Duan,  Wenhui and Xu,  Yong},
  year = {2022},
  month = June,
  pages = {367–377}
}

@article{perdew1996generalized,
  title={Generalized gradient approximation made simple},
  author={Perdew, John P and Burke, Kieron and Ernzerhof, Matthias},
  journal={Physical review letters},
  volume={77},
  number={18},
  pages={3865},
  url = {http://dx.doi.org/10.1103/PhysRevLett.77.3865},
  DOI = {10.1103/physrevlett.77.3865},
  year={1996},
  publisher={APS}
}

@article{Lee2024,
  title={Moir{\'e} polar vortex, flat bands, and Lieb lattice in twisted bilayer BaTiO3},
  volume = {10},
  ISSN = {2375-2548},
  url = {http://dx.doi.org/10.1126/sciadv.adq0293},
  DOI = {10.1126/sciadv.adq0293},
  number = {47},
  journal = {Science Advances},
  publisher = {American Association for the Advancement of Science (AAAS)},
  author = {Lee,  Seungjun and de Sousa,  D. J. P. and Jalan,  Bharat and Low,  Tony},
  year = {2024},
  month = Nov 
}

@article{Mele2010,
  title = {Commensuration and interlayer coherence in twisted bilayer graphene},
  volume = {81},
  ISSN = {1550-235X},
  url = {http://dx.doi.org/10.1103/PhysRevB.81.161405},
  DOI = {10.1103/physrevb.81.161405},
  number = {16},
  journal = {Physical Review B},
  publisher = {American Physical Society (APS)},
  author = {Mele,  E. J.},
  year = {2010},
  month = Apr 
}

@article{Mak2022,
  title = {Semiconductor moiré materials},
  volume = {17},
  ISSN = {1748-3395},
  url = {http://dx.doi.org/10.1038/s41565-022-01165-6},
  DOI = {10.1038/s41565-022-01165-6},
  number = {7},
  journal = {Nature Nanotechnology},
  publisher = {Springer Science and Business Media LLC},
  author = {Mak,  Kin Fai and Shan,  Jie},
  year = {2022},
  month = July,
  pages = {686–695}
}

@article{Carr2020,
  title = {Electronic-structure methods for twisted moiré layers},
  volume = {5},
  ISSN = {2058-8437},
  url = {http://dx.doi.org/10.1038/s41578-020-0214-0},
  DOI = {10.1038/s41578-020-0214-0},
  number = {10},
  journal = {Nature Reviews Materials},
  publisher = {Springer Science and Business Media LLC},
  author = {Carr,  Stephen and Fang,  Shiang and Kaxiras,  Efthimios},
  year = {2020},
  month = July,
  pages = {748–763}
}

@article{Shahed2025,
  title = {Prediction of polarization vortices,  charge modulation,  flat bands,  and moiré magnetism in twisted oxide bilayers},
  volume = {111},
  ISSN = {2469-9969},
  url = {http://dx.doi.org/10.1103/PhysRevB.111.195420},
  DOI = {10.1103/physrevb.111.195420},
  number = {19},
  journal = {Physical Review B},
  publisher = {American Physical Society (APS)},
  author = {Shahed,  Naafis Ahnaf and Samanta,  Kartik and Elekhtiar,  Mohamed and Huang,  Kai and Eom,  Chang-Beom and Rzchowski,  Mark S. and Belashchenko,  Kirill D. and Tsymbal,  Evgeny Y.},
  year = {2025},
  month = May 
}

@article{Zhang2025,
  title = {{Tear-and-stack} twisted {SrTiO$_3$} Moir\'e Superlattices for Precise Interfacial Reconstruction and Polar Topology},
  volume = {38},
  ISSN = {1521-4095},
  url = {http://dx.doi.org/10.1002/adma.202519300},
  DOI = {10.1002/adma.202519300},
  number = {9},
  journal = {Advanced Materials},
  publisher = {Wiley},
  author = {Zhang,  Yingli and Ge,  Jinxin and Su,  Shengyao and Li,  Yuhao and Zhang,  Wenxi and Lyu,  Longji and Song,  Jiahao and Liu,  Yuxin and Lei,  Yihan and Du,  Haopeng and Zhong,  Gaokuo and Huang,  Boyuan and Li,  Jiangyu and Li,  Changjian},
  year = {2025},
  month = Dec 
}

@article{Yang2024,
  title = {Evolution of flat bands in MoSe$_2$/WSe$_2$ moir\'e lattices: A study combining machine learning and band unfolding methods},
  volume = {110},
  ISSN = {2469-9969},
  url = {http://dx.doi.org/10.1103/PhysRevB.110.235410},
  DOI = {10.1103/physrevb.110.235410},
  number = {23},
  journal = {Physical Review B},
  publisher = {American Physical Society (APS)},
  author = {Yang,  Shengguo and Chen,  Jiaxin and Liu,  Chao-Fei and Chen,  Mingxing},
  year = {2024},
  month = Dec 
}

@article{Cao2018,
  title = {Unconventional superconductivity in magic-angle graphene superlattices},
  volume = {556},
  ISSN = {1476-4687},
  url = {http://dx.doi.org/10.1038/nature26160},
  DOI = {10.1038/nature26160},
  number = {7699},
  journal = {Nature},
  publisher = {Springer Science and Business Media LLC},
  author = {Cao,  Yuan and Fatemi,  Valla and Fang,  Shiang and Watanabe,  Kenji and Taniguchi,  Takashi and Kaxiras,  Efthimios and Jarillo-Herrero,  Pablo},
  year = {2018},
  month = Mar,
  pages = {43–50}
}

@article{Bistritzer2011,
  title = {Moiré bands in twisted double-layer graphene},
  volume = {108},
  ISSN = {1091-6490},
  url = {http://dx.doi.org/10.1073/pnas.1108174108},
  DOI = {10.1073/pnas.1108174108},
  number = {30},
  journal = {Proceedings of the National Academy of Sciences},
  publisher = {Proceedings of the National Academy of Sciences},
  author = {Bistritzer,  Rafi and MacDonald,  Allan H.},
  year = {2011},
  month = July,
  pages = {12233–12237}
}

@article{Andrei2020,
  title = {Graphene bilayers with a twist},
  volume = {19},
  ISSN = {1476-4660},
  url = {http://dx.doi.org/10.1038/s41563-020-00840-0},
  DOI = {10.1038/s41563-020-00840-0},
  number = {12},
  journal = {Nature Materials},
  publisher = {Springer Science and Business Media LLC},
  author = {Andrei,  Eva Y. and MacDonald,  Allan H.},
  year = {2020},
  month = Nov,
  pages = {1265–1275}
}

@article{1Cao2018,
  title = {Correlated insulator behaviour at half-filling in magic-angle graphene superlattices},
  volume = {556},
  ISSN = {1476-4687},
  url = {http://dx.doi.org/10.1038/nature26154},
  DOI = {10.1038/nature26154},
  number = {7699},
  journal = {Nature},
  publisher = {Springer Science and Business Media LLC},
  author = {Cao,  Yuan and Fatemi,  Valla and Demir,  Ahmet and Fang,  Shiang and Tomarken,  Spencer L. and Luo,  Jason Y. and Sanchez-Yamagishi,  Javier D. and Watanabe,  Kenji and Taniguchi,  Takashi and Kaxiras,  Efthimios and Ashoori,  Ray C. and Jarillo-Herrero,  Pablo},
  year = {2018},
  month = Mar,
  pages = {80–84}
}

@article{Serlin2020,
  title = {Intrinsic quantized anomalous Hall effect in a moiré heterostructure},
  volume = {367},
  ISSN = {1095-9203},
  url = {http://dx.doi.org/10.1126/science.aay5533},
  DOI = {10.1126/science.aay5533},
  number = {6480},
  journal = {Science},
  publisher = {American Association for the Advancement of Science (AAAS)},
  author = {Serlin,  M. and Tschirhart,  C. L. and Polshyn,  H. and Zhang,  Y. and Zhu,  J. and Watanabe,  K. and Taniguchi,  T. and Balents,  L. and Young,  A. F.},
  year = {2020},
  month = Feb,
  pages = {900–903}
}

@article{Choi2021,
  title = {Correlation-driven topological phases in magic-angle twisted bilayer graphene},
  volume = {589},
  ISSN = {1476-4687},
  url = {http://dx.doi.org/10.1038/s41586-020-03159-7},
  DOI = {10.1038/s41586-020-03159-7},
  number = {7843},
  journal = {Nature},
  publisher = {Springer Science and Business Media LLC},
  author = {Choi,  Youngjoon and Kim,  Hyunjin and Peng,  Yang and Thomson,  Alex and Lewandowski,  Cyprian and Polski,  Robert and Zhang,  Yiran and Arora,  Harpreet Singh and Watanabe,  Kenji and Taniguchi,  Takashi and Alicea,  Jason and Nadj-Perge,  Stevan},
  year = {2021},
  month = Jan,
  pages = {536–541}
}

@article{Nuckolls2020,
  title = {Strongly correlated Chern insulators in magic-angle twisted bilayer graphene},
  volume = {588},
  ISSN = {1476-4687},
  url = {http://dx.doi.org/10.1038/s41586-020-3028-8},
  DOI = {10.1038/s41586-020-3028-8},
  number = {7839},
  journal = {Nature},
  publisher = {Springer Science and Business Media LLC},
  author = {Nuckolls,  Kevin P. and Oh,  Myungchul and Wong,  Dillon and Lian,  Biao and Watanabe,  Kenji and Taniguchi,  Takashi and Bernevig,  B. Andrei and Yazdani,  Ali},
  year = {2020},
  month = Dec,
  pages = {610–615}
}

@article{Polshyn2020,
  title = {Electrical switching of magnetic order in an orbital Chern insulator},
  volume = {588},
  ISSN = {1476-4687},
  url = {http://dx.doi.org/10.1038/s41586-020-2963-8},
  DOI = {10.1038/s41586-020-2963-8},
  number = {7836},
  journal = {Nature},
  publisher = {Springer Science and Business Media LLC},
  author = {Polshyn,  H. and Zhu,  J. and Kumar,  M. A. and Zhang,  Y. and Yang,  F. and Tschirhart,  C. L. and Serlin,  M. and Watanabe,  K. and Taniguchi,  T. and MacDonald,  A. H. and Young,  A. F.},
  year = {2020},
  month = Nov,
  pages = {66–70}
}

@article{Yankowitz2019,
  title = {Tuning superconductivity in twisted bilayer graphene},
  volume = {363},
  ISSN = {1095-9203},
  url = {http://dx.doi.org/10.1126/science.aav1910},
  DOI = {10.1126/science.aav1910},
  number = {6431},
  journal = {Science},
  publisher = {American Association for the Advancement of Science (AAAS)},
  author = {Yankowitz,  Matthew and Chen,  Shaowen and Polshyn,  Hryhoriy and Zhang,  Yuxuan and Watanabe,  K. and Taniguchi,  T. and Graf,  David and Young,  Andrea F. and Dean,  Cory R.},
  year = {2019},
  month = Mar,
  pages = {1059–1064}
}

@article{Hao2021,
  title = {Electric field–tunable superconductivity in alternating-twist magic-angle trilayer graphene},
  volume = {371},
  ISSN = {1095-9203},
  url = {http://dx.doi.org/10.1126/science.abg0399},
  DOI = {10.1126/science.abg0399},
  number = {6534},
  journal = {Science},
  publisher = {American Association for the Advancement of Science (AAAS)},
  author = {Hao,  Zeyu and Zimmerman,  A. M. and Ledwith,  Patrick and Khalaf,  Eslam and Najafabadi,  Danial Haie and Watanabe,  Kenji and Taniguchi,  Takashi and Vishwanath,  Ashvin and Kim,  Philip},
  year = {2021},
  month = Mar,
  pages = {1133–1138}
}

@article{Lee2019,
  title = {Hidden mechanism for embedding the flat bands of Lieb,  kagome,  and checkerboard lattices in other structures},
  volume = {100},
  ISSN = {2469-9969},
  url = {http://dx.doi.org/10.1103/PhysRevB.100.045150},
  DOI = {10.1103/physrevb.100.045150},
  number = {4},
  journal = {Physical Review B},
  publisher = {American Physical Society (APS)},
  author = {Lee,  Chi-Cheng and Fleurence,  Antoine and Yamada-Takamura,  Yukiko and Ozaki,  Taisuke},
  year = {2019},
  month = July 
}

@article{Choi2019,
  title = {Electronic correlations in twisted bilayer graphene near the magic angle},
  volume = {15},
  ISSN = {1745-2481},
  url = {http://dx.doi.org/10.1038/s41567-019-0606-5},
  DOI = {10.1038/s41567-019-0606-5},
  number = {11},
  journal = {Nature Physics},
  publisher = {Springer Science and Business Media LLC},
  author = {Choi,  Youngjoon and Kemmer,  Jeannette and Peng,  Yang and Thomson,  Alex and Arora,  Harpreet and Polski,  Robert and Zhang,  Yiran and Ren,  Hechen and Alicea,  Jason and Refael,  Gil and von Oppen,  Felix and Watanabe,  Kenji and Taniguchi,  Takashi and Nadj-Perge,  Stevan},
  year = {2019},
  month = Aug,
  pages = {1174–1180}
}

@article{Wang2020,
  title = {Correlated electronic phases in twisted bilayer transition metal dichalcogenides},
  volume = {19},
  ISSN = {1476-4660},
  url = {http://dx.doi.org/10.1038/s41563-020-0708-6},
  DOI = {10.1038/s41563-020-0708-6},
  number = {8},
  journal = {Nature Materials},
  publisher = {Springer Science and Business Media LLC},
  author = {Wang,  Lei and Shih,  En-Min and Ghiotto,  Augusto and Xian,  Lede and Rhodes,  Daniel A. and Tan,  Cheng and Claassen,  Martin and Kennes,  Dante M. and Bai,  Yusong and Kim,  Bumho and Watanabe,  Kenji and Taniguchi,  Takashi and Zhu,  Xiaoyang and Hone,  James and Rubio,  Angel and Pasupathy,  Abhay N. and Dean,  Cory R.},
  year = {2020},
  month = June,
  pages = {861–866}
}

@article{Xian2021,
  title = {Realization of nearly dispersionless bands with strong orbital anisotropy from destructive interference in twisted bilayer MoS2},
  volume = {12},
  ISSN = {2041-1723},
  url = {http://dx.doi.org/10.1038/s41467-021-25922-8},
  DOI = {10.1038/s41467-021-25922-8},
  number = {1},
  journal = {Nature Communications},
  publisher = {Springer Science and Business Media LLC},
  author = {Xian,  Lede and Claassen,  Martin and Kiese,  Dominik and Scherer,  Michael M. and Trebst,  Simon and Kennes,  Dante M. and Rubio,  Angel},
  year = {2021},
  month = Sept 
}

@article{Devakul2021,
  title = {Magic in twisted transition metal dichalcogenide bilayers},
  volume = {12},
  ISSN = {2041-1723},
  url = {http://dx.doi.org/10.1038/s41467-021-27042-9},
  DOI = {10.1038/s41467-021-27042-9},
  number = {1},
  journal = {Nature Communications},
  publisher = {Springer Science and Business Media LLC},
  author = {Devakul,  Trithep and Crépel,  Valentin and Zhang,  Yang and Fu,  Liang},
  year = {2021},
  month = Nov 
}

@article{Xian2019,
  title = {Multiflat Bands and Strong Correlations in Twisted Bilayer Boron Nitride: Doping-Induced Correlated Insulator and Superconductor},
  volume = {19},
  ISSN = {1530-6992},
  url = {http://dx.doi.org/10.1021/acs.nanolett.9b00986},
  DOI = {10.1021/acs.nanolett.9b00986},
  number = {8},
  journal = {Nano Letters},
  publisher = {American Chemical Society (ACS)},
  author = {Xian,  Lede and Kennes,  Dante M. and Tancogne-Dejean,  Nicolas and Altarelli,  Massimo and Rubio,  Angel},
  year = {2019},
  month = July,
  pages = {4934–4940}
}

@article{Walet2021,
  title = {Flat bands, strains, and charge distribution in twisted bilayer {h-BN}},
  volume = {103},
  ISSN = {2469-9969},
  url = {http://dx.doi.org/10.1103/PhysRevB.103.125427},
  DOI = {10.1103/physrevb.103.125427},
  number = {12},
  journal = {Physical Review B},
  publisher = {American Physical Society (APS)},
  author = {Walet,  Niels R. and Guinea,  Francisco},
  year = {2021},
  month = Mar 
}

@article{Li2021,
  title = {Extremely flat band in antiferroelectric bilayer $\alpha$-In$_2$Se$_3$ with large twist-angle},
  volume = {23},
  ISSN = {1367-2630},
  url = {http://dx.doi.org/10.1088/1367-2630/ac17b9},
  DOI = {10.1088/1367-2630/ac17b9},
  number = {8},
  journal = {New Journal of Physics},
  publisher = {IOP Publishing},
  author = {Li,  C F and Zhai,  W J and Li,  Y Q and Tang,  Y S and Zhang,  J H and Chen,  P Z and Zhou,  G Z and Cui,  X M and Lin,  L and Yan,  Z B and Huang,  X K and Jiang,  X P and Liu,  J-M},
  year = {2021},
  month = Aug,
  pages = {083019}
}

@article{Xu2020,
  title = {Correlated insulating states at fractional fillings of moiré superlattices},
  volume = {587},
  ISSN = {1476-4687},
  url = {http://dx.doi.org/10.1038/s41586-020-2868-6},
  DOI = {10.1038/s41586-020-2868-6},
  number = {7833},
  journal = {Nature},
  publisher = {Springer Science and Business Media LLC},
  author = {Xu,  Yang and Liu,  Song and Rhodes,  Daniel A. and Watanabe,  Kenji and Taniguchi,  Takashi and Hone,  James and Elser,  Veit and Mak,  Kin Fai and Shan,  Jie},
  year = {2020},
  month = Nov,
  pages = {214–218}
}

@article{Su2022,
  title = {Tuning colour centres at a twisted hexagonal boron nitride interface},
  volume = {21},
  ISSN = {1476-4660},
  url = {http://dx.doi.org/10.1038/s41563-022-01303-4},
  DOI = {10.1038/s41563-022-01303-4},
  number = {8},
  journal = {Nature Materials},
  publisher = {Springer Science and Business Media LLC},
  author = {Su,  Cong and Zhang,  Fang and Kahn,  Salman and Shevitski,  Brian and Jiang,  Jingwei and Dai,  Chunhui and Ungar,  Alex and Park,  Ji-Hoon and Watanabe,  Kenji and Taniguchi,  Takashi and Kong,  Jing and Tang,  Zikang and Zhang,  Wenqing and Wang,  Feng and Crommie,  Michael and Louie,  Steven G. and Aloni,  Shaul and Zettl,  Alex},
  year = {2022},
  month = July,
  pages = {896–902}
}

@article{Liu2014,
  title = {Exotic electronic states in the world of flat bands: From theory to material},
  volume = {23},
  ISSN = {1674-1056},
  url = {http://dx.doi.org/10.1088/1674-1056/23/7/077308},
  DOI = {10.1088/1674-1056/23/7/077308},
  number = {7},
  journal = {Chinese Physics B},
  publisher = {IOP Publishing},
  author = {Liu,  Zheng and Liu,  Feng and Wu,  Yong-Shi},
  year = {2014},
  month = July,
  pages = {077308}
}

@article{Ji2019,
  title = {Freestanding crystalline oxide perovskites down to the monolayer limit},
  volume = {570},
  ISSN = {1476-4687},
  url = {http://dx.doi.org/10.1038/s41586-019-1255-7},
  DOI = {10.1038/s41586-019-1255-7},
  number = {7759},
  journal = {Nature},
  publisher = {Springer Science and Business Media LLC},
  author = {Ji,  Dianxiang and Cai,  Songhua and Paudel,  Tula R. and Sun,  Haoying and Zhang,  Chunchen and Han,  Lu and Wei,  Yifan and Zang,  Yipeng and Gu,  Min and Zhang,  Yi and Gao,  Wenpei and Huyan,  Huaixun and Guo,  Wei and Wu,  Di and Gu,  Zhengbin and Tsymbal,  Evgeny Y. and Wang,  Peng and Nie,  Yuefeng and Pan,  Xiaoqing},
  year = {2019},
  month = June,
  pages = {87–90}
}

@article{Ozaki2004,
  title = {Numerical atomic basis orbitals from H to Kr},
  volume = {69},
  ISSN = {1550-235X},
  url = {http://dx.doi.org/10.1103/PhysRevB.69.195113},
  DOI = {10.1103/physrevb.69.195113},
  number = {19},
  journal = {Physical Review B},
  publisher = {American Physical Society (APS)},
  author = {Ozaki,  T. and Kino,  H.},
  year = {2004},
  month = May 
}

@article{Hwang2012,
  title = {Emergent phenomena at oxide interfaces},
  volume = {11},
  ISSN = {1476-4660},
  url = {http://dx.doi.org/10.1038/nmat3223},
  DOI = {10.1038/nmat3223},
  number = {2},
  journal = {Nature Materials},
  publisher = {Springer Science and Business Media LLC},
  author = {Hwang,  H. Y. and Iwasa,  Y. and Kawasaki,  M. and Keimer,  B. and Nagaosa,  N. and Tokura,  Y.},
  year = {2012},
  month = Jan,
  pages = {103–113}
}

@article{Tokura2000,
  title = {Orbital Physics in Transition-Metal Oxides},
  volume = {288},
  ISSN = {1095-9203},
  url = {http://dx.doi.org/10.1126/science.288.5465.462},
  DOI = {10.1126/science.288.5465.462},
  number = {5465},
  journal = {Science},
  publisher = {American Association for the Advancement of Science (AAAS)},
  author = {Tokura,  Y. and Nagaosa,  N.},
  year = {2000},
  month = Apr,
  pages = {462–468}
}

@article{Chen2024,
  title = {LaAlO$_3$/SrTiO$_3$ Heterointerface: 20 Years and Beyond},
  volume = {10},
  ISSN = {2199-160X},
  url = {http://dx.doi.org/10.1002/aelm.202300730},
  DOI = {10.1002/aelm.202300730},
  number = {3},
  journal = {Advanced Electronic Materials},
  publisher = {Wiley},
  author = {Chen,  Shunfeng and Ning,  Yuanjie and Tang,  Chi Sin and Dai,  Liang and Zeng,  Shengwei and Han,  Kun and Zhou,  Jun and Yang,  Ming and Guo,  Yanqun and Cai,  Chuanbing and Ariando,  Ariando and Wee,  Andrew T. S. and Yin,  Xinmao},
  year = {2024},
  month = Jan 
}

@article{Singh2024,
  title = {Stoichiometric control of electron mobility and 2D superconductivity at LaAlO$_3$-SrTiO$_3$ interfaces},
  volume = {7},
  ISSN = {2399-3650},
  url = {http://dx.doi.org/10.1038/s42005-024-01644-3},
  DOI = {10.1038/s42005-024-01644-3},
  number = {1},
  journal = {Communications Physics},
  publisher = {Springer Science and Business Media LLC},
  author = {Singh,  Gyanendra and Guzman,  Roger and Saïz,  Guilhem and Zhou,  Wu and Gazquez,  Jaume and Masoudinia,  Fereshteh and Winkler,  Dag and Claeson,  Tord and Fraxedas,  Jordi and Bergeal,  Nicolas and Herranz,  Gervasi and Kalaboukhov,  Alexei},
  year = {2024},
  month = May 
}

@article{Gong2024,
  title = {Generalizing deep learning electronic structure calculation to the plane-wave basis},
  volume = {4},
  ISSN = {2662-8457},
  url = {http://dx.doi.org/10.1038/s43588-024-00701-9},
  DOI = {10.1038/s43588-024-00701-9},
  number = {10},
  journal = {Nature Computational Science},
  publisher = {Springer Science and Business Media LLC},
  author = {Gong,  Xiaoxun and Louie,  Steven G. and Duan,  Wenhui and Xu,  Yong},
  year = {2024},
  month = Oct,
  pages = {752–760}
}

@article{Tang2024,
  title = {A deep equivariant neural network approach for efficient hybrid density functional calculations},
  volume = {15},
  ISSN = {2041-1723},
  url = {http://dx.doi.org/10.1038/s41467-024-53028-4},
  DOI = {10.1038/s41467-024-53028-4},
  number = {1},
  journal = {Nature Communications},
  publisher = {Springer Science and Business Media LLC},
  author = {Tang,  Zechen and Li,  He and Lin,  Peize and Gong,  Xiaoxun and Jin,  Gan and He,  Lixin and Jiang,  Hong and Ren,  Xinguo and Duan,  Wenhui and Xu,  Yong},
  year = {2024},
  month = Oct 
}

@article{Li2024,
  title = {Deep-Learning Density Functional Perturbation Theory},
  volume = {132},
  ISSN = {1079-7114},
  url = {http://dx.doi.org/10.1103/PhysRevLett.132.096401},
  DOI = {10.1103/physrevlett.132.096401},
  number = {9},
  journal = {Physical Review Letters},
  publisher = {American Physical Society (APS)},
  author = {Li,  He and Tang,  Zechen and Fu,  Jingheng and Dong,  Wen-Han and Zou,  Nianlong and Gong,  Xiaoxun and Duan,  Wenhui and Xu,  Yong},
  year = {2024},
  month = Feb 
}

@article{Wang2017,
  title = {First-principles calculation of nonlinear optical responses by Wannier interpolation},
  volume = {96},
  ISSN = {2469-9969},
  url = {http://dx.doi.org/10.1103/PhysRevB.96.115147},
  DOI = {10.1103/physrevb.96.115147},
  number = {11},
  journal = {Physical Review B},
  publisher = {American Physical Society (APS)},
  author = {Wang,  Chong and Liu,  Xiaoyu and Kang,  Lei and Gu,  Bing-Lin and Xu,  Yong and Duan,  Wenhui},
  year = {2017},
  month = Sept 
}

@article{Wang2019,
  title = {First-principles calculation of optical responses based on nonorthogonal localized orbitals},
  volume = {21},
  ISSN = {1367-2630},
  url = {http://dx.doi.org/10.1088/1367-2630/ab3c9c},
  DOI = {10.1088/1367-2630/ab3c9c},
  number = {9},
  journal = {New Journal of Physics},
  publisher = {IOP Publishing},
  author = {Wang,  Chong and Zhao,  Sibo and Guo,  Xiaomi and Ren,  Xinguo and Gu,  Bing-Lin and Xu,  Yong and Duan,  Wenhui},
  year = {2019},
  month = Sept,
  pages = {093001}
}

@misc{HopTB,
  author       = {{HopTB Developers}},
  title        = {{HopTB.jl}: a tight-binding package for electronic and optical response calculations},
  year         = {2024},
  howpublished = {\url{https://github.com/HopTB/HopTB.jl}},
  note         = {Version 0.8.2}
}

@article{Slater1954,
  title = {Simplified LCAO Method for the Periodic Potential Problem},
  volume = {94},
  ISSN = {0031-899X},
  url = {http://dx.doi.org/10.1103/PhysRev.94.1498},
  DOI = {10.1103/physrev.94.1498},
  number = {6},
  journal = {Physical Review},
  publisher = {American Physical Society (APS)},
  author = {Slater,  J. C. and Koster,  G. F.},
  year = {1954},
  month = June,
  pages = {1498–1524}
}

@article{Naik2018,
  title = {Ultraflatbands and Shear Solitons in Moiré Patterns of Twisted Bilayer Transition Metal Dichalcogenides},
  volume = {121},
  ISSN = {1079-7114},
  url = {http://dx.doi.org/10.1103/PhysRevLett.121.266401},
  DOI = {10.1103/physrevlett.121.266401},
  number = {26},
  journal = {Physical Review Letters},
  publisher = {American Physical Society (APS)},
  author = {Naik,  Mit H. and Jain,  Manish},
  year = {2018},
  month = Dec 
}

@article{Sipe2000,
  title = {Second-order optical response in semiconductors},
  volume = {61},
  ISSN = {1095-3795},
  url = {http://dx.doi.org/10.1103/PhysRevB.61.5337},
  DOI = {10.1103/physrevb.61.5337},
  number = {8},
  journal = {Physical Review B},
  publisher = {American Physical Society (APS)},
  author = {Sipe,  J. E. and Shkrebtii,  A. I.},
  year = {2000},
  month = Feb,
  pages = {5337–5352}
}

@article{Young2012,
  title = {First Principles Calculation of the Shift Current Photovoltaic Effect in Ferroelectrics},
  volume = {109},
  ISSN = {1079-7114},
  url = {http://dx.doi.org/10.1103/PhysRevLett.109.116601},
  DOI = {10.1103/physrevlett.109.116601},
  number = {11},
  journal = {Physical Review Letters},
  publisher = {American Physical Society (APS)},
  author = {Young,  Steve M. and Rappe,  Andrew M.},
  year = {2012},
  month = Sept 
}

@article{Spanier2016,
  title = {Power conversion efficiency exceeding the Shockley–Queisser limit in a ferroelectric insulator},
  volume = {10},
  ISSN = {1749-4893},
  url = {http://dx.doi.org/10.1038/nphoton.2016.143},
  DOI = {10.1038/nphoton.2016.143},
  number = {9},
  journal = {Nature Photonics},
  publisher = {Springer Science and Business Media LLC},
  author = {Spanier,  Jonathan E. and Fridkin,  Vladimir M. and Rappe,  Andrew M. and Akbashev,  Andrew R. and Polemi,  Alessia and Qi,  Yubo and Gu,  Zongquan and Young,  Steve M. and Hawley,  Christopher J. and Imbrenda,  Dominic and Xiao,  Geoffrey and Bennett-Jackson,  Andrew L. and Johnson,  Craig L.},
  year = {2016},
  month = Aug,
  pages = {611–616}
}

@article{https://doi.org/10.48550/arxiv.2606.06289,
  title   = {Optical Signature of Moir\'e Superlattices Formed by Twisted {SrTiO$_3$} Membranes},
  url     = {https://arxiv.org/abs/2606.06289},
  DOI     = {10.48550/arXiv.2606.06289},
  journal = {arXiv preprint arXiv:2606.06289},
  author  = {Shahriar, T. A. M. Ragib and Murakami, Fumikazu and He, Xing and Koons, Konnor and Li, Xinyan and Kim, Bumseop and Qin, Shihan and Harbola, Varun and Mannhart, Jochen and Han, Yimo and Xu, Ruijuan and Huang, Shengxi and Rappe, Andrew M. and Zhu, Hanyu},
  year    = {2026},
  eprint  = {2606.06289},
  archivePrefix = {arXiv},
  primaryClass  = {cond-mat.mtrl-sci}
}

\end{document}